\documentclass[pra,twocolumn,superscriptaddress,showpacs,floatfix]{revtex4-2}
\usepackage{hyperref}
\usepackage[british]{babel}
\usepackage[utf8]{inputenc}
\usepackage{amsmath}
\usepackage{amssymb}
\usepackage{mathtools}
\usepackage{graphicx}
\usepackage{xcolor}

\newcommand{\bs}[1]{\boldsymbol{#1}} 

\begin{document}
\title{Local-photon model of the momentum of light}
\author{Gabriel Waite}
\affiliation{The School of Physics and Astronomy, University of Leeds, Leeds LS2 9JT, United Kingdom.}
\affiliation{Centre for Quantum Computation and Communication Technology.}
\affiliation{Centre for Quantum Software and Information, School of Computer Science, Faculty of Engineering and Information Technology, University of Technology Sydney, NSW 2007, Australia.}
\author{Daniel Hodgson}
\affiliation{The School of Physics and Astronomy, University of Leeds, Leeds LS2 9JT, United Kingdom.}
\author{Ben Lang}
\affiliation{The School of Physics and Astronomy, University of Nottingham, Nottingham  NG7 2RD, United Kingdom.}
\author{Varghese Alapatt}
\affiliation{Instituto de Ciencia Molecular (ICMol), Universitat de València, Paterna, Spain.}
\author{Almut Beige}
\affiliation{The School of Physics and Astronomy, University of Leeds, Leeds LS2 9JT, United Kingdom.}

\date{\today}

\begin{abstract}
Recently we introduced a local photon approach for modelling the quantised electromagnetic field in position space. Using this approach, we define the momentum of light in this paper as in quantum mechanics as the generator for spatial translation. Afterwards, we analyse the momentum dynamics of photonic wave packets which transition from air into a denser dielectric medium. Our analysis shines new light onto the Abraham-Minkowski controversy which highlights the intricacies involved in the characterisation of the momentum of the electromagnetic field. Although our results align with Minkowski's theory and with the definition of the canonical momentum of light in quantum electrodynamics, there are also some crucial differences.
\end{abstract}

\maketitle
\section{Introduction} \label{sec:1}

In standard electrodynamics, the total energy of the electromagnetic (EM) field in a one-dimensional, non-dispersive and homogeneous dielectric medium with permittivity $\varepsilon$ and permeability $\mu$ is given by 
\begin{eqnarray} \label{energy1}
H_{\text{eng}} &=& {A \over 2} \int_{-\infty}^{\infty}\text{d}x \left[ \varepsilon \, \|\boldsymbol{E}(x)\|^2 + \mu \, \|\boldsymbol{H}(x)\|^2\right] \notag \\
&=& {A \over 2} \int_{-\infty}^{\infty}\text{d}x \left[ {1 \over \varepsilon} \, \|\boldsymbol{D}(x)\|^2 + \frac{1}{\mu} \, \|\boldsymbol{B}(x)\|^2\right] . 
\end{eqnarray}
Here $A$ denotes the area occupied by the light in the $y$-$z$ plane when travelling along the $x$ axis, and $\bs{E}(x)$ and $\bs{B}(x)$ denote the local (real) electric and magnetic field vectors. The vectors $\bs{D}(x) = \varepsilon \, \bs{E}(x)$ and $\bs{H}(x) = \bs{B}(x) /\mu$ denote the (real) displacement and magnetising field vectors respectively. From looking at Eq.~(\ref{energy1}) above, it is not clear whether either $\bs{E}(x)$ and $\bs{H}(x)$ or $\bs{D}(x)$ and $\bs{B}(x)$ are the fundamental field vectors for light in a dielectric medium. It is therefore not surprising that there are different definitions for the momentum of light in classical electrodynamics. 

Currently, there are two main definitions of the momentum of the EM field, both of which date back to the beginning of the 20th century. According to Minkowski \cite{Minkowski}, the momentum of light ${\bs p}$ should be written as
\begin{eqnarray}\label{E3}
p_{\rm Min} \, \hat{\bs{x}} &=& A \int_{- \infty}^\infty {\rm d}x \, \bs{D}(x) \times \bs{B}(x) 
\end{eqnarray}
where $\hat{\bs{x}}$ is a unit vector pointing in the direction of the positive $x$ axis. Abraham \cite{Abraham1, Abraham2}, on the other hand, suggested that ${\bs p}$ should be written as 
\begin{eqnarray}\label{E2}
p_{\rm Ab} \, \hat{\bs{x}} &=& {A \over c_0^2} \int_{- \infty}^\infty {\rm d}x \, \bs{E}(x) \times \bs{H}(x) 
\end{eqnarray}
with $c_0$ denoting the speed of light in air. In the following $c=1/(\varepsilon \mu)^{1/2}$ will denote the speed of light in a dielectric medium whilst $n = c_0/c$ will denote its refractive index. Using this notation one can show that $p_{\rm Min}$ and $p_{\rm Ab}$ differ by a factor of $n^2$, and the controversy over this disparity has become known as the Abraham-Minkowski controversy \cite{Mansuripur,Crenshaw,Crenshaw1,Brevik,Leonhardt,Griffiths,Barnett,Pfeifer,Tulkki,Huang21,Orni2}. It has been pointed out, meanwhile, that classical electrodynamics allows for many other possible definitions of the momentum of light in a dielectric medium \cite{Kinsler}. 

Importantly, the different points of view lead to different predictions when considering light transitioning from air into a denser dielectric medium with $n>1$. For example, Abraham's expression leads to a decrease of the momentum of the incoming light by a factor of $n$ \cite{Abraham1,Abraham2} while Minkowski's expression predicts an increase by a factor of $n$ \cite{Minkowski}. Although these predictions seem inconsistent, strong arguments have been made in favour of each.  The Abraham momentum, for instance, can be derived in an ``Einstein Box" experiment which considers the conservation of the mass-energy centre as light propagates through the dielectric medium \cite{Padgett, Balazs, Baxter}.  Alternatively, Abraham's momentum arises when the mechanical momentum of the medium is separated from the total momentum of the field and dielectric \cite{Gordon, Jones1}.  In this regard, the Abraham momentum is usually associated with the flow of energy, and moreover leaves the electromagnetic energy-momentum tensor fully symmetric.  The Minkowski momentum, on the other hand, takes into account the response \cite{Milonni}, and in the quantum picture can be associated with the de Broglie wavelength of a photon \cite{Padgett}.  

In more recent investigations, it has been proposed that both the Abraham and the Minkowski momentum can be expressed as a consistent part of the same physical theory \cite{Baxter, Griffiths}. In Refs.~\cite{Hinds, Barnett2}, for instance, the Abraham and the Minkowski momentum are associated with the kinetic and the canonical momentum of light respectively; although it has also been claimed that this resolution cannot hold in all inertial reference frames \cite{Wang}. Unfortunately, the experimental tests designed to determine the momentum of light have also not been able to provide a clear resolution to the controversy as observations of both the Abraham \cite{Walker, She} and the Minkowski momentum have been reported. In photon drag \cite{Gibson, Mansuripur2,Tulkki} or radiation pressure experiments \cite{Jones2, Ashkin}, such as in the Jones and Leslie experiment \cite{Jones3} which measured the force on a mirror suspended in a fluid, the Minkowski momentum is usually determined \cite{Brevik1}. New proposals for an experimental resolution, however, may be successful in the future \cite{Chen}.  Nevertheless, often it is possible to justify both forms of the momenta provided that the EM forces on the material are properly accounted for \cite{Webb}.

In this paper we therefore take an alternative approach to identifying the momentum of light inside a homogeneous dielectric medium. To do so we notice that a recently introduced local photon approaches \cite{Franson,Jake,Daniel,Daniel2} allow us to assign not only a state vector $|\psi \rangle$ but also a wave function $\psi(x,t)$ to individual photons. By taking into account the position wave function of a photon, we can identify operators representing the Hamiltonian and momentum of light in the same way as we would in quantum mechanics. According to the Schr\"odinger equation, the Hamiltonian of a quantum system is the generator of time translations. In the following, we therefore define the dynamical Hamiltonian $H_{\rm dyn}$ of light such that 
\begin{eqnarray} \label{E5}
H_{\rm dyn} \, |\psi \rangle &=& {\rm i} \hbar \, {\partial \over \partial t} \, |\psi \rangle \, .
\end{eqnarray}
Similarly, the momentum of a quantum mechanical point particle is the generator for spatial translations. Hence we define the dynamical momentum $ p_{\rm dyn}$ of the quantised EM field such that 
\begin{eqnarray} \label{E6}
 p_{\rm dyn} \, |\psi \rangle &=& - {\rm i} \hbar \, {\partial \over \partial x} \, |\psi \rangle \, .
\end{eqnarray}
Without a wave function $\psi(x,t)$ for individual photons, the spatial derivative on the right-hand side of Eq.~(\ref{E6}) would remain meaningless \cite{Arwa}. For more detailed discussions of previous difficulties with defining single photon wave functions see Refs.~\cite{AliM2006,Birula,Sipe,SmithRaymer, Hawton07}.

\begin{figure}[t]
\includegraphics[width=0.48\textwidth]{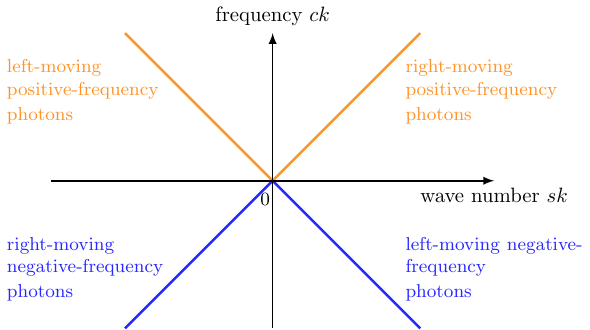}
\caption{As recently shown in Refs.~\cite{Jake,Daniel}, the introduction of a wave function for individual photons requires a doubling of the standard Hilbert space of the quantised EM field. In the extended Hilbert space we label monochromatic plane-wave photons by three quantum numbers: $s$, $k$ and $\lambda$ with $s = \pm 1$, $k \in (-\infty,\infty)$ and $\lambda = {\sf H},{\sf V}$. The frequency and the wave number of these photons are given by $\omega = ck$ and by $sk$ respectively, while $\lambda$ denotes their polarisation. Different from standard theories, we now have photons with positive (orange) and with negative (blue) frequencies.} \label{fig2}
\end{figure}

Proceeding as described in Refs.~\cite{Jake,Daniel,Daniel2} and using the same momentum space photon annihilation and creation operators, $\tilde a_{s \lambda}(k)$ and $\tilde a^\dagger_{s \lambda}(k)$, it can be shown that the dynamical Hamiltonian $H_{\rm dyn}$ is given by 
\begin{eqnarray}
	\label{dynamical Hamiltonian2}
	H_{\text{dyn}} &=& \sum_{s= \pm 1} \sum_{\lambda = {\sf H},{\sf V}} \int_{-\infty}^{\infty}\text{d}k \, \hbar ck \, \tilde a^\dagger_{s \lambda}(k)\, \tilde a_{s \lambda}(k) 
\end{eqnarray}
in the momentum space representation. Here $s$ and $\lambda$ characterise the direction of propagation and the polarisation of monochromatic photons while $k$ varies between plus and minus infinity (cf.~Fig.~\ref{fig2}). A closer look at Eq.~(\ref{dynamical Hamiltonian2}) immediately shows that $H_{\text{dyn}}$ has positive and negative eigenvalues. Our formalism therefore distinguishes between positive- and negative-frequency photons. This applies since for every photonic wave packet of light there is another wave packet that travels in the opposite direction and carries field vectors that also solve Maxwell's equations \cite{book,Ornigotti}. Since the generators of time translations for these two wave packets must differ by a minus sign, a complete description of the quantised EM field requires a Hamiltonian with an equal number of positive {\em and} negative eigenvalues. Only in the case of monochromatic waves with positive frequencies $\omega = ck$ do the above expressions coincide with the standard expression for the energy observable \cite{Bennett}. The importance of including negative frequency photons in the modelling of photonic devices, however, is currently becoming more and more recognised \cite{Dirac, Cook, Cook2, Faccio, AliM2017,Hawton17,Pendry,Hawton23,Hawton24}. 

Since localised photons with a well-defined direction of propagation $s$ travel at the speed of light, it is not surprising that the generator for time translations, $H_{\rm dyn}$, and the generator for spatial translations, $p_{\rm dyn}$, have many similarities. As we shall see below, $p_{\rm dyn}$ equals
\begin{eqnarray}
	\label{dynamical mom2}
	p_{\text{dyn}} &=& \sum_{s= \pm 1} \sum_{\lambda = {\sf H},{\sf V}} \int_{-\infty}^{\infty}\text{d}k \, \hbar sk \, \tilde a^\dagger_{s \lambda}(k)\, \tilde a_{s \lambda}(k) 
\end{eqnarray}
in its momentum space representation. A monochromatic photon with frequency $\omega = ck$ and direction of propagation $s$, therefore, has the momentum $\hbar sk$. For positive $k$, this momentum is the same as the canonical momentum of light \cite{Barnett,Huttner2}. This too is not surprising since quantum electrodynamics usually maps the energy observable of the quantised electromagnetic field onto a harmonic oscillator. Once this link is established, one can identify the canonical momentum of light with the expression for the momentum of a quantum mechanical point particle which does not contain its mass $m$. Moreover, in most situations the sign of $k$ can be ignored when momentum conservation is applied as dielectric interfaces do not convert photons with positive frequencies $k>0$ into photons with negative frequencies $k<0$ and vice versa \cite{Jake}. Nevertheless, in general Eqs.~(\ref{dynamical Hamiltonian2}) and (\ref{dynamical mom2}) should be used.

By definition, the dynamical Hamiltonian $H_{\rm dyn}$ is not only the generator of translations in time but also represents the energy of the quantised EM field \cite{Noether}.  Whilst the eigenvalues of the dynamical Hamiltonian $H_{\rm dyn}$ must necessarily take both positive and negative values \cite{Jake,Daniel}, the eigenvalues of the energy observable $H_{\rm eng}$ usually only take positive values. In this paper, therefore, we assume that
\begin{eqnarray}
	\label{dynamical Hamiltonian78}
	H_{\text{dyn}} &=& \left\{ \begin{array}{ll} - H_{\text{eng}} & ~~{\rm for} ~ k<0 \, , \\ + H_{\text{eng}} & ~~{\rm for} ~ k \ge 0 \, , \end{array} \right. 
\end{eqnarray}
thereby ignoring the energy of the vacuum state, i.e.~the zero point energy of the EM field, which has no dynamics. By taking this into account we show in this paper that the energy observable $H_{\rm eng}$ equals
\begin{eqnarray}
\label{dynamical Hamiltonian76}
H_{\text{eng}} = { A \over 4} \int_{-\infty}^{\infty}\text{d}x \left[ \varepsilon\,\boldsymbol{\cal E}^\dagger(x) \cdot \boldsymbol{\cal E}(x)
+ \frac{1}{\mu}\,\boldsymbol{\cal B}^\dagger(x) \cdot \boldsymbol{\cal B}(x)\right] ~
\end{eqnarray}
with $ \boldsymbol{\cal E}^\dagger(x)$ and $\boldsymbol{\cal B}^\dagger(x)$ denoting the complex electric and magnetic field vector observables respectively. Similarly, we show in this paper that the dynamical momentum $p_{\rm dyn}$ of light can be written in the form
\begin{eqnarray}\label{E12}
p_{\rm dyn} &=& \left\{ \begin{array}{ll} - p & ~ {\rm for} ~ k < 0 \, , \\ + p  & ~ {\rm for} ~ k \ge 0 \end{array} \right.
\end{eqnarray}
with $p$ defined such that
\begin{eqnarray}\label{E11}
p \, \hat{\bs{x}} &=&  {\varepsilon A \over 4} \, \int_{- \infty}^\infty {\rm d}x \, \left[{\boldsymbol{\cal E}}^\dagger (x) \times {\boldsymbol{\cal B}}(x) - {\boldsymbol{\cal B}}^\dagger (x) \times {\boldsymbol{\cal E}}(x) \right] \, . \notag \\
\end{eqnarray}
Comparing this equation with Eq.~(\ref{E3}) shows that $p$ has many similarities with Minkowski's momentum of light \cite{Minkowski}. The only differences are that the real electric and magnetic field vectors $\boldsymbol{E} (x)$ and $\boldsymbol{B} (x)$ have been replaced by an Hermitian combination of their complex counterparts, namely ${\boldsymbol{\cal E}}^\dagger (x) $ and $\boldsymbol{\cal B} (x)$, and a factor $1/4$ has been added.

\begin{figure}[t]
\includegraphics[width=0.48\textwidth]{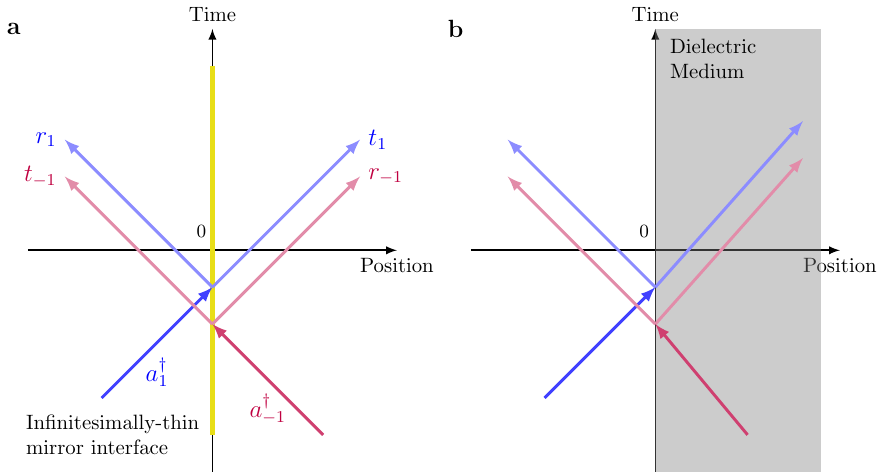}
\caption{({\bf a}) An illustration of light scattering by a partially transparent interface in the $x=0$ plane with the same medium on both sides. Here $r_s$ and $t_s$ denote the reflection and transmission rate of a blip which travels initially in the $s$ direction. As we shall see below, energy conservation and symmetry arguments impose certain conditions onto these rates. ({\bf b}) An illustration of light scattering on the surface of a dielectric medium with a refractive index $n>1$ on the right-hand side of the $x=0$ plane.} \label{fig4}
\end{figure}

In order to compare the energy and the momentum of light before and after transitioning from air into a denser dielectric medium, we shall derive a locally acting mirror Hamiltonian $H_{\rm mir}$ which is a special example of the locally acting Hermitian mirror Hamiltonian in Ref.~\cite{Jake}. The dynamics associated with $H_{\rm mir}$ describe light scattering by a partially transparent mirror surface, conserve photon numbers and can be analysed in a relatively straightforward way. By considering the reflection and transmission rates of light that are consistent with Stokes' relations and Fresnel's coefficients \cite{Hecht}, we find that the energy of any incoming photons is conserved, but there is a change in momentum. When we look only at the eventually transmitted contribution, the increase is by a factor $n$ in agreement with Minkowski's theory \cite{Minkowski}. However, overall, the momentum expectation value $\langle p_{\rm dyn} \rangle$ changes in an unexpected way which is different from previous predictions.

Quantum optics approaches which only consider positive frequency photons (see e.g.~Ref.~\cite{Bennett} and references therein) are widely used in the literature and have been highly successful in explaining experiments. This is not surprising, since single photon experiments usually employ uni-directional monochromatic photons which are much longer than their wavelengths \cite{Kuhn,Kok}. This makes it easy to describe them using momentum space annihilation operators which do not seem to depend on $s$. However, an incomplete description of the electromagnetic field makes it impossible to assign wave functions and position operators to single photons, as required by wave particle duality \cite{Jake,Daniel,Daniel2}. Moreover, as we illustrate in the following, the construction of locally acting mirror Hamiltonians requires locally acting, bosonic field annihilation operators $a_{s \lambda}(x)$ which are the Fourier transforms of the above-mentioned $\tilde a_{s \lambda}(k)$ operators. Due to the locality of physical interactions, the results presented here are likely to provide useful tools for the modelling of complex photonic devices, including dielectric media with dispersion \cite{Raymer2} and space and time varying dielectric media \cite{Pendry,KS,Orni2}.

This paper is structured as follows. Section \ref{sec:2} outlines the classical equations for light propagation in a homogeneous dielectric medium, which are pivotal for obtaining a local quantum theory of the EM field. Section \ref{sec:3} extends our free space model to the experimental setups in Fig.~\ref{fig4} with the same or different media on either side of a partially transparent mirror surface. Afterwards, in Section \ref{sec:4}, we derive expressions for the energy and the dynamical momentum of the quantised EM field thereby confirming Eqs.~(\ref{dynamical Hamiltonian78})-(\ref{E11}). Once this is done, we study the dynamics of their expectation values within the local photon framework and gain new insight into the Abraham-Minkowski controversy. Finally, we summarise our results in Section \ref{sec:5}. 

\section{Local photons in a homogeneous dielectric medium} \label{sec:2}

In this section we review a recently introduced local photon model of the quantised EM field in free space \cite{Jake,Daniel,Daniel2}. To motivate this approach we shall first have a closer look at the classical dynamics of light in free space before quantising the EM field in position space in Section \ref{sec:22}. When transforming the position space operators into momentum space operators in Section \ref{momrep}, we see that the local photon approach requires a doubling of the usual number of photon degrees of freedom that appear in the standard description of the quantised EM field \cite{Bennett}, as illustrated in Fig.~\ref{fig2}. Local photons are therefore not the same as spatial-temporal or temporal modes \cite{Franson,Huttner,Raymer} although there are some similarities. 
  
\subsection{The classical dynamics of photonic wave packets} \label{sec:21}

For simplicity, we restrict ourselves in the following to light propagating along the $x$ axis. In the presence of a homogeneous, non-dispersive dielectric medium, Maxwell's equations tell us that
\begin{eqnarray}\label{eq:MME4}
        \boldsymbol{\nabla} \cdot \boldsymbol{\cal E} (x) = 0 \, , ~~ \boldsymbol{\nabla} \times \boldsymbol{\cal E} (x) = - \frac{\partial}{\partial t} \boldsymbol{\cal B} (x) \, , \notag \\
        \boldsymbol{\nabla} \cdot \boldsymbol{\cal B} (x) = 0 \, , ~~ \boldsymbol{\nabla} \times \boldsymbol{\cal B} (x) = \frac{1}{c^2} \frac{\partial}{\partial t} \boldsymbol{\cal E} (x)\, . 
\end{eqnarray}
The usually considered {\em real} electric and magnetic field vectors $\boldsymbol{E}(x)$ and $\boldsymbol{B}(x)$ are given by the real parts of the complex vectors in Eq.~(\ref{eq:MME4}). After eliminating either $\boldsymbol{\cal E} (x)$ or $\boldsymbol{\cal B} (x)$ in Eq.~(\ref{eq:MME4}) we find the following wave equations:
\begin{eqnarray} \label{eq:MME_WE}
    \left( \frac{\partial^2}{\partial x^2} - \frac{1}{c^2} \frac{\partial^2}{\partial t^2} \right) {\boldsymbol {\cal O}} (x,t) = 0 \, ,
\end{eqnarray}
which act independently on each component of ${\boldsymbol {\cal O}} = \boldsymbol{\cal E}, \boldsymbol{\cal B}$. As pointed out by d'Alembert, it is useful to write the quadratic operator in the above wave equation as the product of two linear operators showing that there are different types of solutions ${\boldsymbol {\cal O}}_{s \lambda} (x,t)$, each satisfying
\begin{eqnarray}\label{eq:classical_eom_diff}
    \left( \frac{\partial}{\partial x}+ \frac{s}{c} \frac{\partial}{\partial t} \right){\boldsymbol {\cal O}}_{s\lambda}(x,t) = 0 
\end{eqnarray}
where $s = \pm 1$. By solving this first-order differential equation one may further show that 
\begin{eqnarray} \label{o15}
    {\boldsymbol {\cal O}}_{s\lambda}(x,t) = {\boldsymbol {\cal O}}_{s\lambda}(x-sct,0), 
\end{eqnarray}
which demonstrates that Maxwell's equations support photonic wave packets with polarisation $\lambda$ that propagate at the speed of light in the direction of either the positive or the negative $x$ axis \cite{book,Ornigotti}. The orientations of the electric and magnetic field vectors ${\boldsymbol {\cal E}}_{s\lambda}(x,t)$ and ${\boldsymbol {\cal B}}_{s\lambda}(x,t)$ depend on both $s$ and $\lambda$. The only difference between light propagation in air and in a dielectric medium is the respective speed at which the light is propagating.

\subsection{The position representation} \label{sec:22}

The discussion in the previous subsection demonstrates that it is possible to decompose wave packets of light not only into monochromatic waves, but also into local building blocks  which travel at the speed of light. In the following we refer to these local quantised building blocks as blips, which stands for {\em bosons localised in position}. As we shall see below, blips are the carriers of electric and magnetic fields, much in a similar way as to how massive objects, e.g.~planets, are carriers of gravitational fields \cite{Jake,Daniel}. Each blip is characterised by a position $x \in (-\infty, \infty)$, a polarisation $\lambda = {\sf H},{\sf V}$ specifying the orientation of the fields carried by the blip and a direction of propagation $s =\pm 1$. 

The corresponding blip annihilation operators $a_{s\lambda}(x)$ must obey the bosonic commutation relation 
\begin{eqnarray}  \label{comm}
\big[ a_{s\lambda} (x), a^\dagger_{s'\lambda'} (x') \big] &=& \delta_{s s'} \, \delta_{\lambda \lambda'} \, \delta(x-x')
\end{eqnarray}
whilst all other commutators return zero. This relation guarantees that the single-blip states 
\begin{eqnarray} \label{st}
|1_{s \lambda} (x) \rangle &=& a^\dagger_{s\lambda}(x) \, |0 \rangle  
\end{eqnarray}
correspond to pairwise orthogonal states when defined at different positions \cite{Jake,Daniel}. 

To find the dynamical Hamiltonian $H_{\rm dyn}$ (Eq.~(\ref{E5})) of the quantised EM field in a dielectric medium, we shall use the associated unitary time evolution operator $U_{\rm dyn}(t,0)$. Transforming the blip annihilation operators $a_{s\lambda}(x)$ into the Heisenberg picture \cite{Loudon} with respect to $t=0$,  we obtain the Heisenberg operators  
\begin{eqnarray}\label{eq:blip_eom2}
    a_{s\lambda}(x,t) &=& U^\dagger_{\rm dyn}(t,0) \, a_{s\lambda}(x) \, U_{\rm dyn}(t,0) 
\end{eqnarray}
with $a_{s\lambda}(x,0) =a_{s\lambda}(x)$. The time- and position-dependent operators $a_{s\lambda}(x,t)$ describe blips with spacetime coordinates $(x,t)$ and polarisation $\lambda$ which travel at the speed of light $c$ in the $s$ direction. Hence, consistency with Eq.~(\ref{o15}) requires that \cite{Jake,Daniel}
\begin{eqnarray}\label{eq:blip_eom}
    a_{s\lambda}(x,t) &=& a_{s\lambda}(x - sct) \, .
\end{eqnarray}
Next we will show that the Hamiltonian 
\begin{eqnarray} \label{eq:dynamical_ham}
H_{\text{dyn}} &=& - {\rm i} \hbar \sum_{s = \pm 1}\sum_{\lambda = {\sf H},{\sf V}} \int_{-\infty}^{\infty}\text{d}x \, sc \, a^\dagger_{s\lambda}(x) \, {\partial \over \partial x} \, a_{s\lambda}(x) \notag \\
\end{eqnarray}
generates these dynamics. Notice that the above $H_{\text{dyn}}$ is the same as the dynamical Hamiltonian which we recently introduced in Refs.~\cite{Jake,Daniel}, but written here in a more compact form.

By taking the time derivative of Eq.~(\ref{eq:blip_eom2}), the $a_{s\lambda}(x,t)$ operators evolve according to the von Neumann equation   
\begin{eqnarray}\label{eq:erhenfest_theorem}
\frac{\partial}{\partial t} a_{s\lambda}(x,t) &=& - \frac{\rm i}{\hbar} \, \left[ a_{s\lambda}(x,t) , H_{\rm dyn} \right] \, .
\end{eqnarray}
Subsequently substituting $H_{\rm dyn}$ in Eq.~(\ref{eq:dynamical_ham}) into this equation and then using Eq.~(\ref{eq:blip_eom}), we may show that
\begin{eqnarray}\label{eq:erhenfest_theoremxxx}
&& \hspace*{-0.8cm} \frac{\partial}{\partial t} a_{s\lambda}(x,t) \notag \\
&=& - sc \, \int_{-\infty}^{\infty}\text{d}x' \, \left[ a_{s\lambda}(x-sct) ,  a^\dagger_{s\lambda}(x') \right] \, {\partial \over \partial x'} \, a_{s\lambda}(x') \notag \\ 
&=& - sc \, \int_{-\infty}^{\infty}\text{d}x' \, \delta(x-x'-sct) \, {\partial \over \partial x'} \, a_{s\lambda}(x') \, .
\end{eqnarray}
This expression can then be simplified with the substitution $\tilde x = x' + sct$ in Eq.~(\ref{eq:erhenfest_theoremxxx}), which leads us to 
\begin{eqnarray}\label{eq:erhenfest_theoremxxx2}
\frac{\partial}{\partial t} a_{s\lambda}(x,t) = - sc \, \int_{-\infty}^{\infty}\text{d}\tilde x \, \delta(x-\tilde x) \, {\partial \over \partial \tilde x} \, a_{s\lambda}(\tilde x,t) 
\end{eqnarray}
and finally to
\begin{eqnarray}\label{eq:blip_eom_diff}
\frac{\partial}{\partial t} a_{s\lambda}(x,t) &=& - sc \,\frac{\partial}{\partial x} a_{s\lambda}(x,t) \, . 
\end{eqnarray}
This means that $H_{\rm dyn}$ is consistent with Eq.~(\ref{eq:blip_eom}) and is therefore the dynamical Hamiltonian of the quantised EM field.  

As shown in Refs.~\cite{Cook, Cook2,Jake,Daniel}, the {\em complex} local electric and magnetic field vectors $\boldsymbol{\cal E}(x)$ and $\boldsymbol{\cal B}(x)$ of photonic wave packets can be written in the form   
\begin{eqnarray}\label{E&B}
        \boldsymbol{\cal E}(x) &=& \sum_{s = \pm 1} R_{s {\sf H}}(x) \, \hat{\boldsymbol{y}} + R_{ s{\sf V}}(x) \, \hat{\boldsymbol{z}} \, ,  \notag \\
        \boldsymbol{\cal B}(x) &=& \sum_{s = \pm1} \, {s \over c} \, \big[ R_{s {\sf H}}(x) \, \hat{\boldsymbol{z}} - R_{s{\sf V}}(x) \, \hat{\boldsymbol{y}} \big]
\end{eqnarray}
where $\hat{\boldsymbol{y}}$ and $\hat{\boldsymbol{z}}$ are unit vectors pointing along the direction of the positive $y$ and $z$ axes respectively. Moreover $R_{s\lambda} (x)$ denotes the non-local annihilation operator
\begin{eqnarray}\label{eq:R_relation_a}
R_{s\lambda} (x) &=& \int_{-\infty}^{\infty} \text{d}x'\, \mathcal{R}(x-x') \, a_{s\lambda}(x') 
\end{eqnarray}
with the distribution $\mathcal{R}(x-x')$ given by \cite{Daniel}
\begin{eqnarray}\label{eq:R_form}
    \mathcal{R}(x-x') &=& - \left(\frac{\hbar c}{4 \pi \varepsilon A} \right)^{1/2} \frac{1}{|x-x'|^{3/2}} \, .
\end{eqnarray}
As ${\cal{R}}(x-x')$ is non-zero for any $x \neq x'$, the complex field operators $\boldsymbol{\cal E}(x)$ and $\boldsymbol{\cal B}(x)$ are the sum of contributions from blips at all points along the $x$ axis. As we shall illustrate in the next subsection, the above choice of $R_{s\lambda} (x)$ guarantees Lorentz covariance \cite{Daniel}. Moreover, notice that the right-hand side of Eq.~(\ref{eq:R_form}) depends on parameters characterising the medium, like the area $A$ that the fields occupy in the $y$-$z$ plane. 

\subsection{The momentum representation} \label{momrep}

For later convenience, and since most readers will be more familiar with it, we conclude this section with a review of the momentum representation of the quantised EM field. When transferring the blip annihilation operators $a_{s\lambda} (x)$ into momentum space, we obtain the bosonic annihilation operators $\tilde{a}_{s\lambda} (k)$ for monochromatic photons. These relate to the $a_{s\lambda}(x) $ operators via a Fourier transform \cite{Jake,Daniel} and
\begin{eqnarray}
	\label{FieldFT1}
	\tilde{a}_{s\lambda}(k) &=& {1 \over (2\pi)^{1/2}} \int_{-\infty}^{\infty} \text{d}x \, {\rm e}^{-{\rm i}skx} \, a_{s\lambda}(x) \, , \notag \\
	a_{s\lambda}(x) &=&  {1 \over (2\pi)^{1/2}} \int_{-\infty}^{\infty}{\text{d}k} \, {\rm e}^{{\rm i}skx} \, \tilde a_{s\lambda}(k) \, .
\end{eqnarray} 
These equations can be used to check that the momentum space annihilation operators $\tilde a_{s\lambda}(k)$ also obey bosonic commutation relations and that
\begin{eqnarray}
\label{comm10}
\left[ \tilde a_{s\lambda}(k), \tilde a^\dagger_{s'\lambda'}(k') \right]  &=& \delta_{s,s'} \, \delta_{\lambda,\lambda'} \, \delta(k-k')
\end{eqnarray}
with $s=\pm 1$, $\lambda = {\sf H}, {\sf V}$ and $k \in (- \infty, \infty)$. Consequently, the single-excitation states $|1_{s\lambda}(k) \rangle = \tilde a^\dagger_{s\lambda}(k) |0 \rangle$ of monochromatic light are pairwise orthogonal. 

To obtain the momentum space representation of the dynamical Hamiltonian $H_{\rm dyn}$ we now substitute Eq.~(\ref{FieldFT1}) into Eq.~(\ref{eq:dynamical_ham}) and perform the $x$ derivative which gives
\begin{eqnarray}
        \label{dynamical Hamiltonian}
        H_{\text{dyn}} &=& {\hbar c \over 2 \pi} \sum_{s = \pm 1}\sum_{\lambda = {\sf H},{\sf V}} \int_{-\infty}^{\infty}\text{d}x \int_{-\infty}^{\infty} {\rm d}k \int_{-\infty}^{\infty} {\rm d} k' \notag \\
        && \times k' \, {\rm e}^{-{\rm i}s(k-k')x} \, \tilde a^\dagger_{s\lambda}(k) \tilde a_{s\lambda}(k') \, .
\end{eqnarray}
Performing the $x$ integration and then integrating over the resulting delta function leads to Eq.~(\ref{dynamical Hamiltonian2}) in which the above Hamiltonian is diagonal. Notice, however, that the main difference compared to the standard representation of the quantised EM field (cf.~e.g.~Ref.~\cite{Bennett}) is a doubling of the usual Hilbert space. We now characterise photons not only by a variable $k \in (-\infty,\infty)$ and a polarisation $\lambda ={\sf H},{\sf V}$, but also by the direction of propagation $s = \pm 1$. As one can see from Eq.~(\ref{dynamical Hamiltonian2}), the frequency $\omega$ of monochromatic photons equals $\omega = c k$ which can be positive and negative. 

Lastly we shall transform the complex electric and magnetic field vectors $ \boldsymbol{\cal E}(x) $ and $ \boldsymbol{\cal B}(x)$ in Eq.~(\ref{E&B}) into their momentum space representations. Before we can do this we must decompose the field observables into their different $(s,\lambda)$ contributions and write them as
\begin{eqnarray}
	\label{sum}
	{\boldsymbol{\cal O}}(x) &=& \sum_{s= \pm 1} \sum_{\lambda = {\sf H},{\sf V}} \boldsymbol{\cal O}_{s\lambda}(x) 
\end{eqnarray}	
with $\boldsymbol{\cal O} = \boldsymbol{\cal E}, \boldsymbol{\cal B}$. By transforming the component $\boldsymbol{\cal O}_{s\lambda}(x)$ into momentum space we obtain the complex electric and magnetic field vectors $\widetilde{\boldsymbol {\cal E}}(k) $ and $\widetilde {\boldsymbol{\cal B}}(k)$ with
\begin{eqnarray}
	\label{sum2}
	\widetilde {\boldsymbol{\cal O}}(k) &=& \sum_{s= \pm 1} \sum_{\lambda = {\sf H},{\sf V}} \widetilde{\boldsymbol{\cal O}}_{s\lambda}(k) 
\end{eqnarray}
and
\begin{eqnarray}
	\label{FieldFT1E&B}
	\widetilde{\boldsymbol{\cal O}}_{s\lambda}(k) &=& {1 \over (2\pi)^{1/2}} \int_{-\infty}^{\infty} \text{d}x \, {\rm e}^{-{\rm i}skx} \, \boldsymbol{\cal O}_{s\lambda}(x) \, , \notag \\
	\boldsymbol{\cal O}_{s\lambda}(x) &=&  {1 \over (2\pi)^{1/2}} \int_{-\infty}^{\infty} \text{d}k \, {\rm e}^{{\rm i}skx} \, \widetilde{\boldsymbol{\cal O}}_{s\lambda}(k) \, ,
\end{eqnarray}	
in analogy to Eq.~(\ref{FieldFT1}). Combining Eqs.~(\ref{E&B}), (\ref{eq:R_relation_a}) and (\ref{FieldFT1}) with the above equations, we can now show, for example, that 
\begin{eqnarray} \label{E&Bfinal40}
\widetilde{\boldsymbol{\cal E}}_{s {\sf H}}(k) 
&=& - {1 \over 2\pi} \left( {\hbar c \over 4 \pi \varepsilon A} \right)^{1/2} \int_{-\infty}^{\infty} \text{d}k' \int_{-\infty}^{\infty} \text{d}x \int_{-\infty}^{\infty} \text{d}x' \notag \\
&& \times {1 \over |x-x'|^{3/2}} \, {\rm e}^{-{\rm i}s(kx-k'x')} \, \tilde{a}_{s\lambda}(k') \, \hat{\boldsymbol{y}} \, .
    \end{eqnarray}
To simplify this expression, we replace the $x'$ integration by an integration over $\xi = x'-x$ which gives
\begin{eqnarray} \label{E&Bfinal41}
\widetilde{\boldsymbol{\cal E}}_{s {\sf H}}(k) 
&=& -  {1 \over 2\pi} \left( {\hbar c \over 4 \pi \varepsilon A} \right)^{1/2} \int_{-\infty}^{\infty} \text{d}k' \int_{-\infty}^{\infty} \text{d}x \int_{-\infty}^{\infty} \text{d}\xi \notag \\
&& \times {1 \over |\xi|^{3/2}} \, {\rm e}^{-{\rm i}s(k-k')x} \, {\rm e}^{{\rm i}sk' \xi} \, \tilde{a}_{s\lambda}(k') \, \hat{\boldsymbol{y}}  \, .
    \end{eqnarray}
Now we can perform the $x$ integration followed by the $k'$ integration to show that
\begin{eqnarray} \label{E&Bfinal}
   \widetilde{\boldsymbol{\cal E}}_{s {\sf H}}(k) &=& \zeta (k) \, \tilde a_{s {\sf H}}(k) \, \hat{\boldsymbol{y}} 
    \end{eqnarray}
where $\zeta (k)$ is the solution to the integral
\begin{eqnarray} \label{zeta}
\zeta (k) &=& -  \left( {\hbar c \over 4 \pi \varepsilon A} \right)^{1/2} \int_{-\infty}^{\infty} \text{d}\xi \, {1 \over |\xi|^{3/2}} \, {\rm e}^{{\rm i}sk \xi} \notag \\
&=& \left( {2 \hbar c \over \varepsilon A} \right)^{1/2} \, |k|^{1/2}  \, . 
\end{eqnarray}
Analogously, one can also show that
\begin{eqnarray}\label{E&Bfinal2}
   \widetilde{\boldsymbol{\cal E}}_{s {\sf V}}(k) &=& \zeta(k) \, \tilde a_{s {\sf V}}(k) \, \hat{\boldsymbol{z}} \, , \notag \\
  -  sc \, \widetilde{\boldsymbol{\cal B}}_{s {\sf V}}(k) &=& \zeta(k) \, \tilde a_{s{\sf V}}(k) \, \hat{\boldsymbol{y}} \, , \notag \\
  sc \, \widetilde{\boldsymbol{\cal B}}_{s {\sf H}}(k) &=& \zeta(k) \, \tilde a_{s{\sf H}}(k) \, \hat{\boldsymbol{z}}\, ,
\end{eqnarray}
in addition to Eq.~(\ref{E&Bfinal}). The operators $ \widetilde{\boldsymbol{\cal E}}_{s \lambda}(k)$ and $ \widetilde{\boldsymbol{\cal B}}_{s \lambda}(k)$ coincide with the momentum space annihilation operators  $\tilde a_{s {\sf V}}(k)$ up to a factor that is proportional to $|k|^{1/2}$. The only difference between air and any other dielectric medium are the values of $\varepsilon$ and $\mu$ and therefore also of $c$.

\section{Local photons in the presence of a partially transparent mirror interface} \label{sec:3}

The presence of an optical element, like a partially transparent mirror, does not restrict the possible shapes, polarisations and directions of propagation that a wave packet of light may have at any given time. As a result, the Hilbert space of the quantised EM field needs to remain the same, even when a mirror interface is placed on the $x$ axis as illustrated in Fig.~\ref{fig4}. The only thing that changes is the dynamics of any incoming wave packets. These now split into transmitted and reflected components when coming into contact with a mirror surface. As we show below, this change in dynamics can be accounted for by adding an interaction term $H_{\rm int}$ to the dynamical Hamiltonian $H_{\rm dyn}$ of the quantised EM field. In the presence of a mirror, the total Hamiltonian now equals \cite{Jake}
\begin{eqnarray} \label{matrix4x}
H_{\rm mir} &=& H_{\rm dyn} + H_{\rm int} \, .
\end{eqnarray}
The purpose of this section is to obtain $H_{\rm int}$ for light scattering by a dielectric medium. 

\subsection{Linear optics beamsplitter transformations} \label{sec32}

First let us have a closer look at how the linear optics community describes light scattering by a beamsplitter, i.e.~by an infinitesimally thin partially transparent mirror, with the same medium on both sides (cf.~Fig.~\ref{fig4}(a)). In the following, we denote the complex reflection and transmission rates by $r_s$ and $t_s$ where $s$ indicates the direction of propagation of the incoming wave packet. Placing the beamsplitter in the $x=0$ plane, the scattering operator $S$ is defined such that \cite{Zeilinger2,Lim}
\begin{eqnarray} \label{matrix3}
\left( \begin{array}{r} S \, a^\dagger_{-1 \lambda}(0) \, S^\dagger \\ S \, a^\dagger_{1 \lambda}(0) \, S^\dagger \end{array} \right) 
&=& U \left( \begin{array}{r} a^\dagger_{-1 \lambda}(0) \\ a^\dagger_{1 \lambda}(0) \end{array} \right) 
\end{eqnarray}
with the transition matrix $U$ given by
\begin{eqnarray} \label{matrix3x}
U &=& \left( \begin{array}{cc}  t_{-1}  &  r_1 \\  r_{-1} &  t_1 \end{array} \right) \, .
\end{eqnarray}
The $a^\dagger_{s \lambda}(0)$ on the right-hand side of Eq.~(\ref{matrix3}) are the creation operators of the incoming blips at the mirror surface while the $S a^\dagger_{s \lambda}(0,t) S^\dagger$ on the left are the creation operators of blips that have already experienced the mirror surface. If there is no absorption in the mirror surface, photon numbers must be conserved. This applies when the transition matrix $U$ is unitary, i.e.~when
 \begin{eqnarray} \label{condi2}
     r_{-1}^*  t_1 +  t_{-1}^* r_1 = 0 \, , ~~ |r_{\pm 1}|^2 + |t_{\pm1}|^2 = 1 \, .
\end{eqnarray}
In classical optics these relations are known as Stokes' relations \cite{Hecht}. 

\subsection{An infinitesimally thin mirror surface with the same medium on both sides}
\label{Sec:mirr1}

Linear optics experiments show that the dynamics of individual photons are unaffected by the presence of other photons: each photon travels independently through the mirror interface \cite{Zeilinger2,Lim}. This means that the dynamics of more complex quantum states of light can be deduced from the dynamics of the single-excitation states of the EM field. Therefore, we focus in the following on these states. In addition, we notice that the time evolution operator $U_{\rm mir} (t,0)$ of the mirror Hamiltonian in Eq.~(\ref{matrix4x}) must evolve the state $|1_{s \lambda} (x) \rangle$ as it would in free space, that is, as
\begin{eqnarray} \label{matrix4}
U_{\rm mir} (t,0) \, |1_{s \lambda} (x) \rangle &=& |1_{s \lambda} (x+sct) \rangle
\end{eqnarray}
if the blip does not reach the $x=0$ plane within the time interval $(0,t)$. If the incoming blip does reach the mirror surface, however, then we require that
\begin{eqnarray} \label{matrix4z2}
        U_{\rm mir} (t,0) \, |1_{s \lambda} (x) \rangle &=& t_s \, |1_{s \lambda}(x +sct) \rangle \notag \\
            && + r_s \, |1_{-s \lambda}(-x -sct) \rangle \, .
\end{eqnarray}
To generate the dynamics described in Eqs.~(\ref{matrix4}) and (\ref{matrix4z2}) $H_{\rm int}$ needs to convert left-moving into right-moving blips at $x=0$ and vice versa. In the remainder of this subsection we show that the Hamiltonian 
\begin{eqnarray} \label{matrix5}
H_{\rm int} &=& \sum_{\lambda = {\sf H}, {\sf V}} \hbar \Omega \, a^\dagger_{-1 \lambda}(0) \, a_{1 \lambda}(0) + {\rm H.c.}
\end{eqnarray}
with a complex mirror coupling constant $\Omega$ has these properties. The above interaction Hamiltonian is Hermitian by construction. Moreover, for symmetry reasons, $\Omega$ cannot depend on $s$ or on $\lambda$. As we shall see below, $\Omega$ depends on $t_s$ and $r_s$ and on the speed of the incoming wave packets.

To derive the time evolution operator $U_{\rm mir}(t,0)$ we denote the free space time evolution operator by $U_{\rm dyn}(t,0)$. In addition, we employ a Dyson series expansion
which is obtained after first moving into the interaction picture with respect to $H_0 = H_{\rm dyn}$ and $t=0$ and then evaluating the time evolution operator in the interaction picture with help of the usual Dyson series expansion. Proceeding in this way and returning into the Schr\"odinger picture, one can show that
\begin{widetext}
\begin{eqnarray}\label{BB3}
        U_{\rm mir}(t,0) &=& U_{\rm dyn}(t,0) - {{\rm i} \over \hbar} \int_0^t {\rm d}t_1 \, U_{\rm dyn}(t,t_1) \, H_{\rm int} \, U_{\rm dyn}(t_1,0) \notag  \\
        && + \left( - {{\rm i} \over \hbar} \right)^2 \int_0^t {\rm d}t_2 \int_0^{t_2} {\rm d}t_1 \, U_{\rm dyn}(t,t_2) H_{\rm int} \, U_{\rm dyn}(t_2,t_1) \, H_{\rm int} \, U_{\rm dyn}(t_1,0)  + \ldots \notag \\
        && + \left( - {{\rm i} \over \hbar} \right)^n \int_0^t {\rm d}t_n \ldots \int_0^{t_2} {\rm d}t_1 \, U_{\rm dyn}(t,t_n) \, H_{\rm int} \, U_{\rm dyn}(t_n,t_{n-1}) \ldots  H_{\rm int} \, U_{\rm dyn}(t_1,0) + \ldots \, .
\end{eqnarray}
\end{widetext}
When applying $U_{\rm mir}(t,0)$ to a single-blip excitation state $ |1_{1 \lambda}(x) \rangle$ that does not arrive at the mirror interface within the time interval $(0,t)$, only the first term in Eq.~(\ref{BB3}) contributes to its dynamics. All higher-order terms return zero since
\begin{eqnarray} \label{BBB2}
&& \hspace{-1cm} H_{\rm int} \, U_{\rm dyn}(t_1,0) \, |1_{1 \lambda}(x) \rangle \notag \\
&=& \hbar \Omega \, |1_{-1 \lambda}(0) \rangle \langle 1_{1 \lambda}(0) |1_{1 \lambda}(x + ct_1) \rangle \notag \\
&=& \hbar \Omega \, |1_{-1 \lambda}(0) \rangle \, \delta (x + ct_1) 
\end{eqnarray}
equals zero in this case. Consequently, the blip does not experience $H_{\rm int}$ and evolves exactly as requested in Eq.~(\ref{matrix4}). Similarly, one can show that
\begin{eqnarray} \label{BBB22}
H_{\rm int} U_{\rm dyn}(t_1,0)|1_{-1 \lambda}(x) \rangle &=& \hbar \Omega^* |1_{1 \lambda}(0) \rangle \, \delta (x - ct_1) \notag \\
\end{eqnarray}
for a blip initially propagating to the left. Thus, an outgoing left-moving blip also evolves as if in free space, consistent with Eq.~(\ref{matrix4}).

Next we have a closer look at the dynamics of a blip initially prepared in $ |1_{s \lambda}(x) \rangle$ and arriving at $x=0$ within $(0,t)$. Now {\em all} terms in Eq.~(\ref{BB3}) contribute to the time evolution of the initial state. For example, using Eq.~(\ref{BBB2}) the first-order contribution gives
\begin{eqnarray}\label{BBB}
        && \hspace*{-0.8cm} - {{\rm i} \over \hbar} \int_0^t {\rm d}t_1 \, U_{\rm dyn}(t,t_1) \, H_{\rm int} \, U_{\rm dyn}(t_1,0) \, |1_{s \lambda}(x) \rangle \notag \\
        &=& - {\rm i} \Omega^{(*)} \int_0^t {\rm d}t_1 \, |1_{-s \lambda}(-sc(t-t_1)) \rangle \, \delta(x+sct_1) ~~~
\end{eqnarray}
since $U_{\rm dyn}(t,t_1) \, |1_{-s \lambda}(0) \rangle = |1_{-s \lambda}(-sc(t-t_1)) \rangle$. Here we take $\Omega^{(*)}$ to denote $\Omega$ when $s=1$ and $\Omega^*$ when $s=-1$.  To perform this time integration we substitute $x_1=-sct_1$ which leads us to
\begin{eqnarray} \label{BBBx}
&& \hspace*{-1cm} - {{\rm i} \over \hbar} \int_0^t {\rm d}t_1 \, U_{\rm dyn}(t,t_1) \, H_{\rm int} \, U_{\rm dyn}(t_1,0) \, |1_{s \lambda}(x) \rangle \notag \\
&=& {{\rm i} \Omega^{(*)} \over sc} \int_0^{-sct} {\rm d}x_1 \, |1_{-s \lambda}(-x_1-sct) \rangle \, \delta(x-x_1) \notag \\
&=& - {{\rm i} \Omega^{(*)} \over c} \, |1_{-s \lambda}(-x -sct) \rangle \, . 
\end{eqnarray}
To calculate the $n$-th-order contribution to the time evolution of $|1_{s \lambda}(x) \rangle$ with $n\ge2$ we proceed in analogy to  Eqs.~(\ref{BBB2})-(\ref{BBBx}) and show that
\begin{widetext}
\begin{eqnarray}\label{new}
     && \hspace*{-0.8cm} \left( - {{\rm i} \over \hbar} \right)^n \int_0^t {\rm d}t_n \ldots  \int_0^{t_2} {\rm d}t_1 \, U_{\rm dyn}(t,t_n) \, H_{\rm int} \, U_{\rm dyn}(t_n,t_{n-1}) \ldots H_{\rm int} \, U_{\rm dyn}(t_1,0)  \, |1_{s \lambda}(x) \rangle \notag \\
            &=& \left( - {\rm i} A(\Omega)\right)^n \int_0^t {\rm d}t_n \ldots \int_0^{t_2} {\rm d}t_1 \, |1_{\pm s \lambda}(\pm sc (t - t_n)) \rangle \, \delta( sc(t_n-t_{n-1})) \ldots \delta(sc(t_2-t_1)) \, \delta(-sct_1-x) \, . 
\end{eqnarray}
Here $A(\Omega)^n = |\Omega|^n$ for even $n$ and $A(\Omega)^n = |\Omega|^{n-1}\Omega^{(*)}$ for odd $n$. The delta functions in this equation account for the fact that blips experience $H_{\rm int}$ only when positioned at $x=0$; at all other positions they are not in contact with the mirror interface. Above, the plus sign applies when $n$ is even and the negative sign applies when $n$ is odd. After substituting $x_i = -sct_i$ with $i=1,\ldots,n$ in order to replace all time integrations by integrations in space we see that
\begin{eqnarray} \label{BBB6}
&& \hspace*{-1cm} \left( - {{\rm i} \over \hbar} \right)^n \int_0^t {\rm d}t_n \ldots \int_0^{t_2} {\rm d}t_1 \, U_{\rm dyn}(t,t_n) \, H_{\rm int} \, U_{\rm dyn}(t_n,t_{n-1}) \ldots H_{\rm int} \, U_{\rm dyn}(t_1,0)  \, |1_{s \lambda}(x) \rangle \notag \\
&=& \left({{\rm i} A(\Omega) \over sc} \right)^n \int_0^{-sct} {\rm d}x_n \ldots \int_0^{x_2} {\rm d}x_1 \, |1_{\pm s \lambda}(\pm(sc t+x_n)) \rangle \, \delta (x_n-x_{n-1}) \ldots \delta(x_2-x_1) \, \delta(x_1-x) \notag \\
&=& 2 \left(- {{\rm i} A(\Omega) \over 2c} \right)^n  |1_{\pm s \lambda}( \pm (sc t+x)) \rangle \, .
\end{eqnarray}
The above calculation takes into account that all integrations\textemdash with the exception of one\textemdash cover only half a $\delta$-function. Using Eq.~(\ref{BB3}) and combining all higher-order terms, we therefore find that 
\begin{eqnarray}\label{BBB7}
        U_{\rm mir}(t,0) \, |1_{1 \lambda}(x) \rangle 
        &=& \left[ 1 + 2 \sum_{n=1}^\infty \left(- {{\rm i} |\Omega| \over 2c} \right)^{2n} \right] |1_{1 \lambda}(x + ct) \rangle
        - {{\rm i} \Omega \over c} \sum_{n=0}^\infty \left(- {{\rm i} |\Omega| \over 2c} \right)^{2n} \, |1_{-1 \lambda}(-x - ct) \rangle \, , \notag \\
        U_{\rm mir}(t,0) \, |1_{-1 \lambda}(x) \rangle 
        &=& \left[ 1 + 2 \sum_{n=1}^\infty \left(- {{\rm i} |\Omega| \over 2c} \right)^{2n} \right] |1_{1 \lambda}(x - ct) \rangle
        - {{\rm i} \Omega^* \over c} \sum_{n=0}^\infty \left(- {{\rm i} |\Omega| \over 2c} \right)^{2n} \, |1_{1 \lambda}(-x + ct) \rangle \, .
\end{eqnarray}
\end{widetext}
This equation is of the same form as Eq.~(\ref{matrix4z2}). In the case of reflection, the original blip is replaced by its mirror image which seems to come from the opposite side of the interface. A transmitted blip, however, evolves as it would in the absence of the mirror. 

For $|\Omega| < 2c$ we can perform the above summations and calculate the reflection and transmission rates $r_s$ and $t_s$ of the partially transparent mirror interface in Fig.~\ref{fig4}(a). Doing so we find that 
\begin{eqnarray} \label{BB6}
t_{\pm 1} &=& \frac{1 - (|\Omega|/2c)^2}{1+ (|\Omega|/2c)^2} \, .
\end{eqnarray}
This  expression is real and can assume any value between 0 and 1. For $\Omega =0$ we get $t_{\pm1} = 1$ as one would expect. In addition, both transmission rates become zero when $|\Omega|$ tends to its maximum value of $2c$.  The complex reflection rates $r_{\pm 1}$ are given by
\begin{eqnarray} \label{BB6xx}
r_{-1} = - \frac{{\rm i} \Omega^*/c}{1+ (|\Omega|/2c)^2} \, , ~~
r_1 = - r_{-1}^* \, .
\end{eqnarray}
The phases of these rates depends upon the phase of the complex mirror coupling constant $\Omega$. 

As mentioned already above, photons do not interact with each other\textemdash they only interfere. To show that this is indeed the case, suppose $|0 \rangle$ denotes the vacuum state of the quantised EM field. Then one can show that 
\begin{eqnarray}\label{CC1}
       && \hspace*{-1cm} U_{\rm mir}(t,0) \, [a^\dagger_{s \lambda}(x)]^n \, |0 \rangle \notag \\ 
       &=& \left[ U_{\rm mir}(t,0) \, a^\dagger_{s \lambda}(x) \, U^\dagger_{\rm mir}(t,0) \right]^n |0 \rangle
\end{eqnarray}
which takes into account that $U^\dagger_{\rm mir}(t,0) \, |0 \rangle = |0 \rangle$ and $U_{\rm mir}(t,0) U^\dagger_{\rm mir}(t,0) = 1$. When substituting Eq.~(\ref{matrix4z2}) into Eq.~(\ref{CC1}) we therefore find that 
\begin{eqnarray}\label{62}
      && \hspace*{-1cm} U_{\rm mir}(t,0) \, [a^\dagger_{s \lambda}(x)]^n \, |0 \rangle \notag \\ 
      &=& \left[ t_s \, a^\dagger_{s \lambda}(x+sct) + r_s \, a^\dagger_{-s \lambda}(-x-sct) \right]^n |0 \rangle 
\end{eqnarray}
which describes the independent scattering of all incoming blips. To show that the mirror Hamiltonian $H_{\rm mir}$ also produces the expected dynamics for any possible initial state $|\psi(0) \rangle$, we should write $|\psi(0) \rangle$ as a function of creation operators $[a^\dagger_{s\lambda} (x)]^n$ applied to $|0 \rangle$ where $n$ is an integer. Fortunately this is always possible \cite{Zeilinger2,Lim} and hence $H_{\rm int}$ in Eq.~(\ref{matrix5}) is indeed the interaction Hamiltonian of a partially transparent mirror surface.

\subsection{An infinitesimally thin mirror surface with a different medium on either side} 
\label{Sec:scattering}

Next we have a closer look at the case where the mirror surface appears as a result of placing a dielectric medium along the positive $x$-axis, as is illustrated in Fig.~\ref{fig4}(b). Again, the total Hamiltonian $H_{\rm mir}$ of the quantised EM field is the sum of two terms, $H_{\rm dyn}$ and $H_{\rm int}$ (cf.~Eq.~(\ref{matrix4x})). As before, the dynamical Hamiltonian $H_{\rm dyn}$ describes the propagation of light in the absence of the interface. The only difference between the situations depicted in Figs.~{\ref{fig4}(a)} and {\ref{fig4}(b)} is that light now travels at different speeds on either side of $x=0$. To account for this, we replace Eq.~(\ref{eq:dynamical_ham}) in the following by
\begin{eqnarray}\label{eq:dynamical_ham2prev}
        H_{\text{dyn}} &=& - {\rm i} \hbar \sum_{s=\pm1} \sum_{\lambda = {\sf H}, {\sf V}} 
        \int_{-\infty}^0 \text{d}x \, sc_0 \, a^\dagger_{s\lambda}(x) \, {\partial \over \partial x} \, a_{s\lambda}(x) \notag \\
        && - {\rm i} \hbar \sum_{s=\pm1} \sum_{\lambda = {\sf H}, {\sf V}} 
        \int_0^{\infty} \text{d}x \, sc \, a^\dagger_{s\lambda}(x) \, {\partial \over \partial x} \, a_{s\lambda}(x)  \notag \\
\end{eqnarray}
with $c_0 = n c$. It is tempting to assume that the interaction term\textemdash which accounts for the presence of a mirror surface\textemdash remains the same as in Eq.~(\ref{matrix5}). Unfortunately, this is not the case. From Eqs.~(\ref{BB6}) and (\ref{BB6xx}), we see that the complex coupling constant $\Omega$ depends on how fast the light approaches the mirror surface and this speed is now no longer the same on both sides.
 
Nevertheless, is it possible to describe the situation in Fig.~\ref{fig4}(b) using a Hermitian locally acting mirror Hamiltonian. All we need to do is to map the situation which we consider here onto the situation which we considered in the previous subsection. Suppose the  annihilation operators $b_{s \lambda}(x)$ are defined such that
\begin{eqnarray} \label{trafo}
b_{s \lambda}(x) &=& \left\{ \begin{array}{cl} a_{s \lambda}(x) & ~~ {\rm for} ~~ x \le 0 \, , \\ 
a_{s \lambda}(x/n)/\sqrt{n} & ~~ {\rm for} ~~ x>0 \, . \end{array} \right.
\end{eqnarray}
Using Eq.~(\ref{comm}), one can easily check that these operators too obey bosonic commutation relations: 
\begin{eqnarray}  \label{comm10}
\big[ b_{s\lambda} (x), b^\dagger_{s'\lambda'} (x') \big] &=& \delta_{s s'} \, \delta_{\lambda \lambda'} \, \delta(x-x') \, .
\end{eqnarray}
Most importantly, the dynamical Hamiltonian $H_{\rm dyn}$ in Eq.~(\ref{eq:dynamical_ham2prev}) can now be written as 
\begin{eqnarray} \label{eq:dynbee}
H_{\text{dyn}} &=& - {\rm i} \hbar \sum_{s=\pm1} \sum_{\lambda = {\sf H}, {\sf V}} 
\int_{-\infty}^{\infty} \text{d}x \, sc_0 \, b^\dagger_{s\lambda}(x) \, {\partial \over \partial x} \, b_{s\lambda}(x)  \, . \notag \\
\end{eqnarray}
When modelled by the $b_{s\lambda} (x)$ operators, light seems to approach the mirror surface at $x=0$ at the same speed, as in the situation in Fig.~\ref{fig4}(a).  

In addition, we notice that the photon number operator $N$ does not depend on whether we count photons using the $a_{s \lambda}(x)$ or the $b_{s \lambda}(x)$ operators since
\begin{eqnarray} \label{numbers}
N &=& \sum_{s=\pm1} \sum_{\lambda = {\sf H}, {\sf V}} \int_{-\infty}^{\infty} \text{d}x \, a^\dagger_{s\lambda}(x) a_{s\lambda}(x) \notag \\
&=& \sum_{s=\pm1} \sum_{\lambda = {\sf H}, {\sf V}} \int_{-\infty}^{\infty} \text{d}x \, b^\dagger_{s\lambda}(x) b_{s\lambda}(x)
\end{eqnarray}
by construction. As the calculations from the previous subsection have shown, the mirror interaction Hamiltonian $H_{\rm int}$, written now as
\begin{eqnarray} \label{matrix6}
H_{\rm int} &=& \sum_{\lambda = {\sf H}, {\sf V}} \hbar \Omega \, b^\dagger_{-1 \lambda}(0) \, b_{1 \lambda}(0) + {\rm H.c.}
\end{eqnarray}
in analogy to Eq.~(\ref{matrix5}), combined with the dynamical Hamiltonian in Eq.~(\ref{eq:dynbee}), conserves photon numbers. To model light scattering into and out of a homogeneous dielectric medium, we should therefore use $H_{\rm dyn}$ and $H_{\rm int}$ in Eqs.~(\ref{eq:dynbee}) and (\ref{matrix6}), but with the physical states of any incoming wave packets of light generated by the $a_{s \lambda}(x) $ operators in Eq.~(\ref{trafo}). 

Suppose that at $t=0$ a single blip has initially been placed at a position $x<0$ while the EM field is in its vacuum state everywhere else. In addition, we assume that the blip travels to the right and reaches the $x=0$ plane within the time interval $(0,t)$. Repeating the calculations from the previous subsection for the initial state 
\begin{eqnarray}\label{EE70}
|\psi_{\rm in}(0) \rangle = a^\dagger_{1 \lambda} (x) |0 \rangle = b^\dagger_{1 \lambda} (x) |0 \rangle \, , 
\end{eqnarray}
but now with the speed of light $c$ and all $a$ operators replaced by $c_0$ and by the $b$ operators respectively, one can show that the state vector $|\psi_{\rm out}(t) \rangle = U_{\rm mir} (t,0) |\psi_{\rm in}(0) \rangle$ of the EM field at time $t$ equals
\begin{eqnarray}\label{EE71}
        |\psi_{\rm out}(t) \rangle &=& t_1 \, b^\dagger_{1 \lambda}(x+c_0t) |0 \rangle + r_1 \, b^\dagger_{-1 \lambda}(-x-c_0t) |0 \rangle \notag \\
\end{eqnarray}
with the photon transmission and reflection rates, $t_1$ and $r_1$, of the $b$ operators are given by 
\begin{eqnarray} \label{EE71b}
t_1 = \frac{1- (|\Omega|/2c_0)^2}{1+ (|\Omega|/2c_0)^2} \, , ~~
r_1 = - \frac{{\rm i} \Omega/c_0}{1+ (|\Omega|/2c_0)^2} \, .
\end{eqnarray}
After replacing the above $b$ operators in Eq.~(\ref{EE71}) by $a$ operators, this equation leads us to
\begin{eqnarray}\label{EE72}
        |\psi_{\rm out}(t) \rangle &=& \left( t_1/\sqrt{n} \right) \, |1_{1 \lambda}((x+c_0t)/n) \rangle \notag \\
        && + r_1 \, |1_{-1 \lambda}(-x-c_0t) \rangle 
\end{eqnarray}
with $ |1_{s \lambda} (x) \rangle = a^\dagger_{s \lambda} (x) |0\rangle$. The factors $1/\sqrt{n}$ and $1/n$ in this equation come from Eq.~(\ref{trafo}) and ensure that photon numbers are conserved. The blip density of the incoming light changes accordingly upon interaction with the mirror surface.

Next we consider a single left-moving blip on the right-hand side of the $x=0$ plane which transitions from the dielectric medium into air while there is vacuum everywhere else. In this case, the initial state $|\psi_{\rm in}(0) \rangle$ of the EM field equals 
\begin{eqnarray} \label{EE73}
   |\psi_{\rm in}(0) \rangle = a^\dagger_{-1 \lambda} (x) |0 \rangle = \sqrt{n} \, b^\dagger_{-1 \lambda} (nx) |0 \rangle  
\end{eqnarray}
with $x>0$. Proceeding as above, we now find that the mirror Hamiltonian $H_{\rm mir}$ evolves this state into 
\begin{eqnarray}\label{EE74}
           |\psi_{\rm out}(t) \rangle &=& \sqrt{n} \, t_{-1} \, b^\dagger_{-1 \lambda}(nx-c_0t) |0 \rangle \notag \\
            && + \sqrt{n} \, r_{-1} \, b^\dagger_{1 \lambda}(-nx+c_0t) |0 \rangle 
\end{eqnarray}
if the blip reaches the mirror interface within the time interval $(0,t)$. The transmission and reflection rates in this equation are now given by
\begin{eqnarray} \label{EE71c}
t_{-1} = t_1 \, , ~~
r_{-1} = - r_1^* 
\end{eqnarray}
with $t_1$ and $r_1$ given in Eq.~(\ref{EE71b}). By again making use of Eq.~(\ref{trafo}), we now find that 
\begin{eqnarray}\label{EE75}
        |\psi_{\rm out}(t) \rangle &=& \sqrt{n} \, t_{-1} \, |1_{-1 \lambda}(nx - c_0t) \rangle \notag \\
        && + r_{-1} \, |1_{1 \lambda}(-x+c_0t/n) \rangle 
\end{eqnarray}
for sufficiently large times $t$. Looking at Eqs.~(\ref{EE72}) and (\ref{EE75}), one might get the impression that the dynamics of incoming wave packets are no longer unitary. However, both Hamiltonians, $H_{\rm dyn}$ and $H_{\rm int}$ in Eqs.~(\ref{eq:dynbee}) and (\ref{matrix6}), are Hermitian. The mirror Hamiltonian $H_{\rm mir}$ in Eq.~(\ref{matrix4x}) is therefore also Hermitian.

\section{The energy and the momentum dynamics of the quantised EM field} \label{sec:4}

Unfortunately, the analysis in the previous section does not reveal how the transmission and reflection rates $r_s$ and $t_s$ depends on the refractive index $n$ of the dielectric medium.  This applies since so far we have only exploited one conservation law, namely photon number conservation. In order to identify the dependence of $r_s$ and $t_s$ on the refractive index $n$ of the dielectric medium, an additional constraint is needed. In this section we therefore impose continuity of the electric field vector on the mirror surface and derive the Fresnel coefficients of classical optics \cite{Hecht}. In addition, we obtain explicit expressions for the energy and the momentum of the quantised EM field and confirm the validity of Eqs.~(\ref{dynamical Hamiltonian78})-(\ref{E11}) in the Introduction. When studying the dynamics of the expectation values of both observables for wave packets which transition from air into a homogenous dielectric medium, we find that energy is always conserved but the momentum of light changes. If we only consider the transmitted contribution, the dynamical momentum seems to increase by a factor $n$ in agreement with Minkowski \cite{Minkowski}, since the wave number $k$ increases by a factor $n$. However, overall, the expectation value $\langle p_{\rm dyn} \rangle$ changes in an unexpected way which is different from previous predictions.

\subsection{Energy and momentum observables} \label{sec:4A}

From quantum optics experiments, we know that a single atom generates exactly one photon upon emission \cite{Kuhn}. Since the energy of the emitted photon is the same as the energy now missing from the atom, a monochromatic photon with frequency $\omega = c k$ must have the energy $\hbar |\omega| = \hbar c |k|$. Hence the energy observable $H_{\rm eng}$ of the quantised EM field needs to be of the form
\begin{eqnarray}
	\label{dynamical Hamiltonian74}
	H_{\text{eng}} &=& \sum_{s= \pm 1} \sum_{\lambda = {\sf H},{\sf V}} \int_{-\infty}^{\infty}\text{d}k \, \hbar c |k| \, \tilde{a}^\dagger_{s\lambda}(k) \tilde{a}_{s \lambda}(k) \, . ~~
\end{eqnarray}
This observable is different from the energy observable that we obtained in our earlier papers \cite{Jake,Daniel} after applying the correspondence principle to the usual expression for the energy of the EM in classical electrodynamics. Instead, the above operator has been constructed such that its eigenvectors have the expected energy eigenvalues while unphysical cross terms are avoided. Moreover, comparing $H_{\rm eng}$ in Eq.~(\ref{dynamical Hamiltonian74}) with the dynamical Hamiltonian $H_{\rm dyn}$ in Eq.~(\ref{dynamical Hamiltonian2}), we notice that $H_{\rm eng}$ and $H_{\rm dyn}$ share their eigenstates, but some of their eigenvalues differ by a minus sign, as was pointed out in Eq.~(\ref{dynamical Hamiltonian78}). In other words, the dynamical Hamiltonian $H_{\rm dyn}$ is indeed a measure for the energy of the quantised EM field \cite{Noether}. The sign difference is important, since it ensures that left- and right-moving wave packets do not have the same dynamics, even when they have the same energy.

To be able to write $H_{\rm eng}$ in Eq.~(\ref{dynamical Hamiltonian74}) as a function of the complex electric and magnetic field observables ${\bs {\cal E}}(x)$ and ${\bs {\cal B}}(x)$, we now use Eqs.~(\ref{sum}) and (\ref{sum2}) to show that
\begin{eqnarray}
	\label{dyn70}
	&& \hspace*{-1cm} \int_{-\infty}^{\infty}\text{d}x \, {\boldsymbol{\cal O}}^\dagger (x) \cdot {\boldsymbol{\cal O}}(x) \notag \\
	&=& {1 \over 2 \pi} \sum_{s,s' = \pm 1} \sum_{\lambda,\lambda' = {\sf H},{\sf V}} \int_{-\infty}^{\infty}\text{d}x \int_{-\infty}^{\infty} {\rm d}k \int_{-\infty}^{\infty} {\rm d}k' \notag \\
	&& \times {\rm e}^{-{\rm i}s(k-k')x} \, {\rm e}^{-{\rm i}(s-s')k' x} \,  \widetilde{\boldsymbol{\cal O}}^\dagger_{s \lambda}(k) \cdot \widetilde{\boldsymbol{\cal O}}_{s' \lambda'}(k') ~~
\end{eqnarray}
for $ {\boldsymbol{\cal O}} =  {\boldsymbol{\cal E}}, {\boldsymbol{\cal B}}$. For $s=s'$, the $x$ integration yields a delta function in $k-k'$ which can be used to further simplify the above operator. For $s \neq s'$ this is not possible.  After taking into account contributions from both the electric and magnetic fields in Eq.~(\ref{dynamical Hamiltonian76}), however, the $s\neq s'$ terms cancel and do not contribute to the total energy. Hence, we find that
\begin{eqnarray}
	\label{dyn71}
	&& \hspace*{-1cm} \int_{-\infty}^{\infty} \text{d}x \, {\boldsymbol{\cal O}}^\dagger (x) \cdot {\boldsymbol{\cal O}}(x) \notag \\
	&=& \sum_{s = \pm 1} \sum_{\lambda,\lambda' = {\sf H},{\sf V}} \int_{-\infty}^{\infty} {\rm d}k \, 
	\widetilde{\boldsymbol{\cal O}}^\dagger_{s \lambda}(k) \cdot \widetilde{\boldsymbol{\cal O}}_{s \lambda'}(k) \, .
\end{eqnarray}
Using Eqs.~(\ref{E&Bfinal})-(\ref{E&Bfinal2}), one can therefore show that $H_{\rm eng}$ in Eq.~(\ref{dynamical Hamiltonian74}) can indeed be written as in Eq.~(\ref{dynamical Hamiltonian76}) in the Introduction. For monochromatic waves with positive frequencies, the complex electric and magnetic field vectors contribute equally to the energy and Eq.~(\ref{dynamical Hamiltonian76}) coincides on average with the energy observable which we usually associate with the quantised EM field \cite{Bennett}. In general, however, both expressions are not the same and Eq.~(\ref{dynamical Hamiltonian76}) should be used.

In quantum mechanics, the total momentum of a collection of particles is the generator for spatial displacements. In the following, we take an analogous approach and define the free-space momentum of photonic wave packets $p_{\text{dyn}}$ as was proposed in Eq.~(\ref{E6}). Considering light with a well-defined direction of propagation $s$ and taking into account that it travels at constant speed, it can be seen from Eq.~(\ref{eq:blip_eom_diff}) that the position and the time derivatives of a state vector $|\psi \rangle$ differ by only a constant factor $sc$. Hence, using the short-hand notation 
\begin{eqnarray}
H_{\rm dyn} = \sum_{s=\pm 1} H_{\rm dyn}(s) \, , ~~
p_{\rm dyn} = \sum_{s=\pm 1} p_{\rm dyn}(s) \, ,
\end{eqnarray} 
Eqs.~(\ref{E5}) and (\ref{E6}) imply that  
\begin{eqnarray}
\label{dynamicalmomentum}
p_{\text{dyn}}(s) &=& (s/c) \, H_{\text{dyn}}(s) \, .
\end{eqnarray} 
Combining this observation with Eq.~(\ref{eq:dynamical_ham}) yields the dynamical momentum
\begin{eqnarray} \label{eq:dynamical_mom}
p_{\text{dyn}} = - {\rm i} \hbar \sum_{s = \pm 1}\sum_{\lambda = {\sf H},{\sf V}} \int_{-\infty}^{\infty}\text{d}x \, a^\dagger_{s\lambda}(x) \, {\partial \over \partial x} \, a_{s\lambda}(x) ~~
\end{eqnarray}
in the position representation and Eq.~(\ref{dynamical mom2}) in the momentum representation. Like $H_{\rm dyn}$, the dynamical momentum $p_{\rm dyn}$ has an equal number of positive and negative eigenvalues. Nevertheless, unlike the energy, $p_{\rm dyn}$ has a different sign for state vectors describing wave packets that travel in different directions. This is illustrated in Fig.~\ref{fig2} which shows that the frequency of monochromatic photons equals $\omega = ck$ while their wave number is given by $sk$ \cite{Jake}.

To identify the dependence of the dynamical momentum $p_{\rm dyn}$ on the complex electric and magnetic field vectors $\boldsymbol{\cal E}(x) $ and $\boldsymbol{\cal B}(x)$, we now use Eqs.~(\ref{sum}) and (\ref{FieldFT1E&B}) to show that 
\begin{eqnarray}\label{newdef}
        && \hspace*{-0.8cm} \int_{-\infty}^{\infty}\text{d}x \, \boldsymbol{\cal E}^\dagger(x) \times \boldsymbol{\cal B}(x) \notag \\
        &=& {1 \over 2 \pi} \sum_{s,s' = \pm 1} \sum_{\lambda,\lambda' = {\sf H},{\sf V}} \int_{-\infty}^{\infty}\text{d}x \int_{-\infty}^{\infty} {\rm d}k \int_{-\infty}^{\infty} {\rm d}k'  \notag  \\
        && \times {\rm e}^{-{\rm i}s(k-k')x} \, {\rm e}^{-{\rm i}(s-s')k' x} \, \widetilde{\boldsymbol{\cal E}}^\dagger_{s \lambda}(k) \times \widetilde{\boldsymbol{\cal B}}_{s' \lambda'}(k') ~~
 \end{eqnarray}
 in momentum space. Again, terms with $\lambda \neq \lambda'$ do not contribute to the vector product in the above equation because of the orientations of the $\widetilde{\boldsymbol{\cal E}}_{s \lambda}(k) $ and $\widetilde{\boldsymbol{\cal B}}_{s \lambda}(k) $ vectors (cf.~Eqs.~(\ref{E&Bfinal}) and (\ref{E&Bfinal2})). Moreover, as in the previous subsection, the $x$ integration yields a delta function in $k-k'$ when $s=s'$; but when $s \neq s'$ this is not the case. By taking into account all contributions in Eq.~(\ref{E11}), however, we find that  
\begin{eqnarray} \label{Hamiltonian71b}
        && \hspace*{-0.8cm} \int_{-\infty}^{\infty} {\rm d} x \; \boldsymbol{\cal E}^\dagger(x) \times \boldsymbol{\cal B}(x) + {\rm H.c.} \notag \\
        &=& \sum_{s = \pm 1} \sum_{\lambda = {\sf H},{\sf V}} \int_{-\infty}^{\infty} {\rm d}k \; \widetilde{\boldsymbol{\cal E}}^\dagger_{s \lambda}(k) \times \widetilde{\boldsymbol{\cal B}}_{s \lambda}(k) + {\rm H.c.}
\end{eqnarray}
Substituting Eqs.~(\ref{E&Bfinal})-(\ref{E&Bfinal2}) into this equation, we therefore find that
\begin{eqnarray}\label{Hamiltonian72}
        && \hspace*{-0.8cm} \int_{-\infty}^{\infty}\text{d}x \; \left[\boldsymbol{\cal E}^\dagger(x) \times \boldsymbol{\cal B}(x) - \boldsymbol{\cal B}^\dagger(x) \times \boldsymbol{\cal E}(x)\right] \notag \\
        &=& {4 \over \varepsilon A} \sum_{s= \pm 1} \sum_{\lambda = {\sf H},{\sf V}} \int_{-\infty}^{\infty}\text{d}k \, \hbar s |k| \, \tilde{a}^\dagger_{s\lambda}(k) \tilde{a}_{s \lambda}(k) \, \hat{\boldsymbol{x}} \, . ~~~
\end{eqnarray}
The eigenvalues of the above vector operator are real and positive for left- and real and negative for right-moving wave packets. When comparing the above expression with the dynamical momentum $p_{\rm dyn}$ in Eq.~(\ref{dynamical mom2}), we can verify that Eqs.~(\ref{E12}) and (\ref{E11}) are satisfied. 

\subsection{The effect of the dielectric medium on expectation values} \label{sec4.2}

Next we have a closer look at the situation in Fig.~\ref{fig4}(b) and study how energy and momentum expectation values change when light transitions from air into a dielectric medium. For simplicity and since energy and momentum are both additive, we consider only a single photon with a given polarisation $\lambda$. Assuming that the photon approaches the mirror surface from both sides, we write its initial state vector $|\psi_{\rm in} (0) \rangle$ as
\begin{eqnarray} \label{oma} 
|\psi_{\rm in} (0) \rangle &=& \sum_{s = \pm 1} \int_{-\infty}^\infty {\rm d}x \, \psi_{s \lambda} (x) \, |1_{s \lambda} (x) \rangle ~~~
\end{eqnarray}
which is normalised when 
\begin{eqnarray} \label{oma2} 
\sum_{s = \pm 1} \int_{-\infty}^\infty {\rm d}x \, \left| \psi_{s \lambda} (x) \right|^2 &=& 1 \, . 
\end{eqnarray}
Suppose that the incoming light is initially far away from the mirror surface, but is eventually scattered according to Eqs.~(\ref{EE72}) and (\ref{EE75}). Then the state vector of the outgoing photon equals 
\begin{eqnarray} \label{oma3} 
|\psi_{\rm out} (t) \rangle &=& \int_{-\infty}^\infty {\rm d}x \, \left[ \sqrt{n} \, t_{-1} \, \psi_{-1 \lambda} (x) \, |1_{-1 \lambda}(nx - c_0t) \rangle \right. \notag \\
&& + r_{-1} \, \psi_{-1 \lambda} (x) \, |1_{1 \lambda}(-x+c_0t/n) \rangle \notag \\
&& + (t_1 /\sqrt{n}) \, \psi_{1 \lambda} (x) \, |1_{1 \lambda}((x+c_0t)/n) \rangle \notag \\
&& \left. + r_1 \, \psi_{1 \lambda} (x) \, |1_{-1 \lambda}(-x-c_0t) \rangle \right] 
\end{eqnarray}
after a sufficiently large time $t$. In principle, we now know everything we need to know to analyse the dynamics of expectation values. However, the following calculations are best done in the momentum basis. 

By transforming the above states with the help of Eq.~(\ref{FieldFT1}) into the momentum basis we find that 
\begin{eqnarray} 
\label{mominstate}
|\psi_{\rm in} (0) \rangle &=& \sum_{s = \pm 1} \int_{-\infty}^\infty {\rm d}k \, \widetilde \psi_{s \lambda} (k) \, |1_{s \lambda} (k) \rangle ~~~
\end{eqnarray}
with the normalised wave function $\tilde \psi_{s \lambda} (k)$ given by
\begin{eqnarray} \label{oma4}
\widetilde{\psi}_{s\lambda}(k) &=& {1 \over (2\pi)^{1/2}} \int_{-\infty}^{\infty} \text{d}x \, {\rm e}^{-{\rm i}skx} \, \psi_{s\lambda}(x) \, .
\end{eqnarray} 
Moreover, the state of the scattered photon equals
\begin{eqnarray}  
|\psi_{\rm out} (t) \rangle &=& {1 \over (2 \pi)^{1/2}} \int_{-\infty}^\infty {\rm d}k  \int_{-\infty}^\infty {\rm d}x \notag \\
&& \times \left[ \sqrt{n} \, t_{-1} \, \psi_{-1 \lambda} (x) \, {\rm e}^{{\rm i}k(nx-c_0t)} \, |1_{-1 \lambda}(k) \rangle \right. \notag \\
&& + r_{-1} \, \psi_{-1 \lambda} (x) \, {\rm e}^{{\rm i}k (x-c_0t/n)} \, |1_{1 \lambda}(k) \rangle \notag \\
&& + (t_1 /\sqrt{n}) \, \psi_{1 \lambda} (x) \, {\rm e}^{-{\rm i}k (x+c_0t)/n} \, |1_{1 \lambda}(k) \rangle ~~ \notag \\
&& \left. + r_1 \, \psi_{1 \lambda} (x) \, {\rm e}^{-{\rm i}k (x+c_0t)} \, |1_{-1 \lambda}(k) \rangle \right] 
\end{eqnarray}
in the momentum representation. After substituting Eq.~(\ref{oma4}) into this equation, $|\psi_{\rm out} (t) \rangle$ simplifies to
\begin{eqnarray} \label{oma5} 
|\psi_{\rm out} (t) \rangle &=& \int_{-\infty}^\infty {\rm d}k
\left[ \sqrt{n} \, t_{-1} \, \widetilde \psi_{-1 \lambda} (nk) \, {\rm e}^{-{\rm i}c_0kt} \, |1_{-1 \lambda}(k) \rangle \right. \notag \\
&& + r_{-1} \, \widetilde \psi_{-1 \lambda} (k) \, {\rm e}^{-{\rm i}c_0kt/n} \, |1_{1 \lambda}(k) \rangle \notag \\
&& + (t_1 /\sqrt{n}) \, \widetilde \psi_{1 \lambda} (k/n) \, {\rm e}^{-{\rm i}c_0kt/n} \, |1_{1 \lambda}(k) \rangle \notag \\
&& \left. + r_1 \, \widetilde \psi_{1 \lambda} (k) \, {\rm e}^{-{\rm i}c_0 kt} \, |1_{-1 \lambda}(k) \rangle \right] \, .
\end{eqnarray}
The above equation shows that scattering by a dielectric medium does not change the sign of $k$.  When the photon transitions from air into the medium, $k$ is multiplied by the refractive index $n$, exactly as originally predicted by Minkowski \cite{Minkowski}. When propagating out of the medium, $k$ is divided by $n$.

At times when the photonic wave packet is not in contact with the mirror surface, i.e.~well before and well after the scattering has taken place, the energy and the momentum observables of the quantised EM field can be approximated by $H_{\rm eng}$ and $p_{\rm dyn}$ in Eqs.~(\ref{dynamical Hamiltonian2})-(\ref{dynamical Hamiltonian78}). Initially we shall consider a photon approaching the medium from the left, through air. Using Eq.~(\ref{mominstate}), we find that the energy and momentum expectation values $\langle H_{\rm eng}^{\rm in} \rangle$ and $\langle p_{\rm dyn}^{\rm in} \rangle$ of the incoming wave packet equal 
\begin{eqnarray}\label{oma6}
         \langle H_{\rm eng}^{\rm in} \rangle &=& \int_{-\infty}^{\infty}\text{d}k \, \hbar c_0 |k| \left| \widetilde \psi_{1 \lambda}(k) \right|^2 \, , \notag\\
         \langle p_{\rm dyn}^{\rm in} \rangle &=& \int_{-\infty}^{\infty}\text{d}k \, \hbar k \left| \widetilde \psi_{1 \lambda}(k) \right|^2
\end{eqnarray}
in this case. Since the reflected and transmitted contributions to the outgoing wave packet travel through air and the dielectric medium respectively, and since $c=c_0/n$, Eq.~(\ref{oma5}) implies that the expectation values $\langle H_{\rm eng}^{\rm out} \rangle$ and $\langle p_{\rm dyn}^{\rm out} \rangle$ of the outgoing light equal
\begin{eqnarray}\label{oma66}
         \langle H_{\rm eng}^{\rm out} \rangle &=& \int_{-\infty}^{\infty}\text{d}k \, \hbar c_0 |k| \notag \\
         && \hspace*{-0.8cm} \times \left[ {|t_1|^2 \over n^2} \left| \widetilde \psi_{1 \lambda}(k/n) \right|^2 + |r_1|^2 \left| \widetilde \psi_{1 \lambda}(k) \right|^2 \right] \, , \notag \\
         \langle p_{\rm dyn}^{\rm out} \rangle &=& \int_{-\infty}^{\infty}\text{d}k \, \hbar k \notag \\
         && \hspace*{-0.8cm} \times \left[ {|t_1|^2 \over n} \left| \widetilde \psi_{1 \lambda}(k/n) \right|^2 - |r_1|^2 \left| \widetilde \psi_{1 \lambda}(k) \right|^2 \right] ~~~ 
\end{eqnarray}
which leads us to
\begin{eqnarray}\label{oma7}
         \langle H_{\rm eng}^{\rm out} \rangle &=&  \left( |t_1|^2 + |r_1|^2 \right) \langle H_{\rm eng}^{\rm in} \rangle \, , \notag \\
         \langle p_{\rm dyn}^{\rm out} \rangle &=& \left( n \, |t_1|^2 - |r_1|^2 \right) \langle p_{\rm dyn}^{\rm in} \rangle \, .
\end{eqnarray}
Considering only the eventually transmitted contribution, we see that the momentum of light increases by a factor $n$. This observation agrees with Minkowski's prediction \cite{Minkowski} and shines some new light onto the so-called Abraham-Minkowski controversy.

For completeness, we also have a closer look at an incoming wave packet that approaches the $x=0$ plane from the right and transitions from the medium into air. Now
\begin{eqnarray}\label{oma12}
         \langle H_{\rm eng}^{\rm in} \rangle &=& {1 \over n} \int_{-\infty}^{\infty}\text{d}k \, \hbar c_0 |k| \left| \widetilde \psi_{-1 \lambda}(k) \right|^2 \, , \notag \\
         \langle p_{\rm dyn}^{\rm in} \rangle &=& - \int_{-\infty}^{\infty}\text{d}k \, \hbar k \left| \widetilde \psi_{-1 \lambda}(k) \right|^2 \, .
\end{eqnarray}
Again using Eq.~(\ref{oma5}) and proceeding as above, one can show that the energy and momentum expectation values of the outgoing light are now given by
\begin{eqnarray}\label{oma66}
         \langle H_{\rm eng}^{\rm out} \rangle &=& \int_{-\infty}^{\infty}\text{d}k \, \hbar c_0 |k| \notag \\
         && \hspace*{-0.8cm} \times \left[ n |t_{-1}|^2 \left| \widetilde \psi_{-1 \lambda}(nk) \right|^2 + {|r_{-1}|^2 \over n} \left| \widetilde \psi_{-1 \lambda}(k) \right|^2 \right] \, , \notag \\
         \langle p_{\rm dyn}^{\rm out} \rangle &=& - \int_{-\infty}^{\infty}\text{d}k \, \hbar k \notag \\
         && \hspace*{-0.8cm}  \times \left[ n |t_{-1}|^2 \left| \widetilde \psi_{-1 \lambda}(nk) \right|^2 - |r_{-1}|^2 \left| \widetilde \psi_{-1 \lambda}(k) \right|^2 \right] ~~~~
\end{eqnarray}
which implies that
\begin{eqnarray}\label{oma13}
        \langle H_{\rm eng}^{\rm out} \rangle &=& \left( |t_{-1}|^2 + |r_{-1}|^2 \right) \langle H_{\rm eng}^{\rm in} \rangle \, , \notag \\
         \langle p_{\rm dyn}^{\rm out} \rangle &=& \left( |t_{-1}|^2 / n - |r_{-1}|^2 \right) \langle p_{\rm dyn}^{\rm in} \rangle \, .
\end{eqnarray}
Since the single-photon reflection and transmission rates $r_s$ and $t_s$ satisfy Stokes' relations in Eq.~(\ref{condi2}), we immediately see from the above equations that the energy of light is conserved. This is not unexpected since the incoming photon evolves unitarily according to a Schr\"odinger equation. However, without knowing the dependence of the rates $|t_s|^2$ and $|r_s|^2$ on $n$, we cannot decide whether or not the same applies to the momentum of light. 

\subsection{Momentum dynamics from the Fresnel coefficients of classical optics} \label{sec4.3}

The Fresnel coefficients $R_s$ and $T_s$ with $s=\pm 1$ of classical electrodynamics are the reflection and transmission rates of the electric field amplitudes for light scattering at the interface between two dielectric materials \cite{Hecht}. They emerge from the assumption that the electric field is continuous across the mirror surface, which is only possible if electric field amplitudes accumulate a minus sign upon reflection by an optically denser medium. For the relatively simple case of normal incidence, we therefore know that 
\begin{eqnarray}
&& R_{-1} = {n-1 \over n+1} \, , ~~ R_1 = - {n-1 \over n+1} \, , ~~ \notag \\
&& T_{-1} = {2n \over n+1} \, , ~~ T_1 = {2 \over n+1} 
\end{eqnarray}
for the experimental setup shown in Fig.~\ref{fig4}(b). Eq.~(\ref{oma3}) moreover tells us how local field amplitudes $\psi_{s \lambda}(x)$ change upon reflection. Since the $\psi_{s \lambda}(x)$ change by the same factor as the local electric field amplitudes change upon reflection, we can deduce that 
\begin{eqnarray}
r_s &=& R_s  
\end{eqnarray}
for $s=\pm 1$. This means that the blip reflection rates $r_s$ are both real and $r_1$ is always negative while $r_{-1}$ must be positive.

As discussed in the beginning of Section \ref{sec:3}, Stokes' relations establish a relation between the mirror transmission and reflection rates $t_s$ and $r_s$. Taking these into account, we therefore also know that 
\begin{eqnarray}
t_{-1} = T_{-1}/ \sqrt{n} \, , ~~ t_1 = \sqrt{n} \, T_1 \, .
\end{eqnarray}
Otherwise, Eq.~(\ref{condi2}) does not hold. As expected, the rates $t_{-1}$ and $t_1$ are therefore the same and real and given by
\begin{eqnarray} \label{Fresnel_results}
t_{\pm1} = \frac{2 \sqrt{n}}{1 + n} \, .
\end{eqnarray}
Moreover, using Eq.~(\ref{EE71b}), one can now show that the value of the coupling constant $\Omega$ that leads to these coefficients is given by
\begin{eqnarray}
\label{Omega_n}
\Omega &=& -2{\rm i}c_0 \, \frac{\sqrt{n}-1}{\sqrt{n}+1} \, .
\end{eqnarray}
For light scattering by an optically denser medium, $\Omega$ is purely imaginary provided we fix the phase relation between electric field vectors and blip annihilation operators as we did in Section \ref{sec:2}, for example. 

As mentioned already above, the energy of an incoming wave packet is conserved after scattering from either side of the boundary. By substituting the above transmission and reflection rates into Eqs.~(\ref{oma7}) and (\ref{oma13}), we  moreover find that the momentum of light transitioning from air into a dielectric medium with $n>1$ changes upon scattering into
\begin{eqnarray}
\label{mom_out1}
\langle p_{\rm dyn}^{\rm out} \rangle &=& \left( 1+2r_{-1} \right) \langle p_{\rm dyn}^{\rm in} \rangle \, .
\end{eqnarray}
This equation implies a momentum increase by a factor  
\begin{eqnarray}
\label{mom_out1b}
1+2r_{-1} &=& {3n-1 \over n+1} \, .
\end{eqnarray}
For light approaching the boundary of the dielectric medium from the right, the momentum decreases for $n>1$ and 
\begin{eqnarray}
\label{mom_out2}
\langle p_{\rm dyn}^{\rm out} \rangle &=& \left( 1-2r_{-1} \right) \langle p_{\rm dyn}^{\rm in} \rangle 
\end{eqnarray}
with
\begin{eqnarray}
\label{last}
1- 2r_{-1} &=& {3-n \over n+1} \, .
\end{eqnarray}
These momentum changes are not surprising since the experimental setup in Fig.~\ref{fig4}(b) does not have the translational invariance required for momentum conservation \cite{Noether}. Not only does the mirror mark a position along the $x$ axis as special, the speed of light decreases inside the medium. The dynamical momentum must therefore change, even in the presence of reflection, in order to propagate the light at the correct speed.  

\section{Conclusions}\label{sec:5}

This paper introduces a general expression for the energy $H_{\rm eng}$ of light in a homogeneous dielectric medium and identifies its dynamical momentum $p_{\rm dyn}$ which is the generator for spatial translations of photonic wave packets. Eqs.~(\ref{dynamical Hamiltonian76}) and (\ref{E11}) show the dependence of $H_{\rm eng}$ and $p_{\rm dyn}$ on the complex electric and magnetic field vector observables $\boldsymbol{\cal E}(x) $ and $\boldsymbol{\cal B} (x) $.  However, our equations do not coincide with the usual expressions for the energy and the momentum of light \cite{Bennett,Kinsler}. They only become the same when only positive-frequency photons are considered and time averages are taken.  As one can see from Eqs.~(\ref{dynamical Hamiltonian2}) and (\ref{dynamical mom2}), in general, monochromatic photons with wave number $sk$ and energy $\hbar c |k|$ have the momentum $\hbar s k$. For light propagation in free space, the energy and the momentum of light are always conserved as one would expect from the spatial and time translational symmetries that are present in this case \cite{Noether}.

Another main result of this paper is the locally acting Hermitian interaction Hamiltonian $H_{\rm mir}$ in~Eq.~(\ref{matrix5}) which can be used to describe the dynamics of the quantised EM field in the presence of a partially reflecting mirror interface. Our analysis emphasises that light scattering is a surface effect. Moreover, we find that the mirror coupling constant $\Omega$ depends on the speed at which light approaches and on the optical properties of the mirror interface. In the case of light scattering by a homogeneous dielectric medium, we obtain an explicit expression for $\Omega$ as a function of its refractive index $n$ (cf.~Eq.~(\ref{Omega_n})). When scattering light with a mirror of finite thickness, there are at least two reflecting surfaces and the effective overall transmission and reflection rates of the mirror need to be derived by either adding terms to the system Hamiltonian or by calculating these rates as in classical optics when analysing light propagation through a so-called Fabry-Perot cavity \cite{Hecht}.

Notice that the experimental setup in Fig.~\ref{fig4}(b) does not have the spatial invariance required for momentum conservation \cite{Noether}. Hence when light transitions from air into a dielectric medium with a higher refractive index, only energy is conserved. In particular, whilst photonic wave packets lose momentum when exiting the medium, wave packets transitioning from air into the medium increase in momentum. This observation can be explained by the slower speed of light inside the medium. Inside the medium, the effective separation between points along the $x$-axis is decreased and large distances now appear very short. As a result, the dynamical momentum must increase in the medium in order the translate wave packets across the contracted distances. It is therefore not surprising that, even in the presence of reflection, the total momentum of an incoming wave packet is not conserved.     
 
By constructing a dynamical momentum operator analogous to the momentum operator of quantum mechanics and studying the dynamics of its expectation values, our analysis shines some new light onto the Abraham-Minkowski controversy. Our observed increase in momentum when light transitions from air into a dielectric medium agrees with Minkowski's treatment of the situation \cite{Minkowski}. Our analysis is based on a local photon approach \cite{Jake,Daniel,Daniel2} and therefore allows for a more straightforward and intuitive description of light scattering than approaches based on infinitely spread out monochromatic waves which are difficult to normalise. Furthermore, our paper provide new tools, for example, for the modelling of light scattering in space and time varying dielectric media and in dielectric media with dispersion, which is still a subject of ongoing research \cite{Raymer,Pendry,KS,Orni2}. \\[0.5cm]
\noindent {\em Acknowledgement.} The authors would like to thank Arwa Bukhari, David Jennings, Axel Kuhn and Robert Purdy for helpful discussions. G. W. was supported by a scholarship from the Sydney Quantum Academy and also supported by the ARC Centre of Excellence for Quantum Computation and Communication Technology (CQC2T), project number CE170100012. D.H. acknowledges financial support from the UK Engineering and Physical Sciences Research Council EPSRC [grant number EP/W524372/1].

\end{document}